\documentclass[a4paper,english,british]{IEEEtran}
\usepackage[T1]{fontenc}
\usepackage[latin9]{inputenc}
\usepackage{color}
\usepackage{amsmath}
\usepackage{amssymb}
\usepackage{graphicx}
\usepackage[numbers]{natbib}

\makeatletter

\providecommand{\tabularnewline}{\\}

\usepackage{amsfonts}\@ifundefined{definecolor}
 {\usepackage{color}}{}
\usepackage{array}\usepackage{hhline}
\makeatletter\newcommand{\ps@Standard}{
  \renewcommand\@oddhead{}
  \renewcommand\@evenhead{}
  \renewcommand\@oddfoot{}
  \renewcommand\@evenfoot{}
  \renewcommand\thepage{\arabic{page}}
}\makeatother
\makeatother

\usepackage{babel}
\begin{document}

\title{Position Reconstruction in a Dual Phase Xenon Scintillation Detector\author{V.N.~Solovov, V.A.~Belov, D.Yu.~Akimov, H.M.~Araújo, E.J.~Barnes,  A.A.~Burenkov, V.~Chepel, A.~Currie, L.~DeViveiros, B.~Edwards, C.~Ghag, A.~Hollingsworth, M.~Horn, G.E.~Kalmus, A.S.~Kobyakin, A.G.~Kovalenko, V.N.~Lebedenko, A.~Lindote, M.I.~Lopes, R.~Lüscher, P.~Majewski, A.St\,J.~Murphy, F.~Neves, S.M.~Paling, J.~Pinto da Cunha, R.~Preece, J.J.~Quenby, L.~Reichhart,  P.R.~Scovell, C.~Silva,  N.J.T.~Smith, P.F.~Smith, V.N.~Stekhanov, T.J.~Sumner, C.~Thorne and R.J.~Walker% 
\thanks{Corresponding author is V.N. Solovov, solovov@coimbra.lip.pt}%
\thanks{E.J.~Barnes, C.~Ghag, A.~Hollingsworth, A.St\,J.~Murphy,    L.~Reichhart  and P.R.~Scovell are with School of Physics \& Astronomy, University of Edinburgh, UK}% 
\thanks{H.M.~Araújo, A.~Currie, M.~Horn, V.N.~Lebedenko, J.J.~Quenby, T.J.~Sumner, C.~Thorne and R.J.~Walker are with High Energy Physics group, Blackett Laboratory, Imperial  College London, UK}% 
\thanks{D.Yu.~Akimov, V.A.~Belov, A.A.~Burenkov, A.S.~Kobyakin, A.G.~Kovalenko and V.N.~Stekhanov are with Institute for Theoretical and Experimental Physics, Moscow, Russia}%
\thanks{V.~Chepel, L.~DeViveiros, A.~Lindote, M.I.~Lopes, F.~Neves, J.~Pinto da Cunha, C.~Silva and V.N.~Solovov are with LIP--Coimbra \& Department of Physics of the University of  Coimbra, Portugal}%
\thanks{B.~Edwards, G.E.~Kalmus, R.~Lüscher, P.~Majewski, S.M.~Paling, R.~Preece, N.J.T.~Smith and P.F.~Smith are with Particle Physics Department, STFC Rutherford Appleton  Laboratory, Chilton, UK}%
}}
\maketitle
\begin{abstract}
We studied the application of statistical reconstruction algorithms,
namely maximum likelihood and least squares methods, to the problem
of event reconstruction in a dual phase liquid xenon detector. An
iterative method was developed for \textit{in-situ} reconstruction
of the PMT light response functions from calibration data taken with
an uncollimated $\gamma$\nobreakdash-ray source. Using the techniques
described, the performance of the ZEPLIN-III dark matter detector
was studied for 122~keV $\gamma$\nobreakdash-rays. For the inner
part of the detector ($R$$<$100~mm), spatial resolutions of 13~mm
and 1.6~mm FWHM were measured in the horizontal plane for primary
and secondary scintillation, respectively. An energy resolution of
8.1\% FWHM was achieved at that energy. The possibility of using this
technique for improving performance and reducing cost of scintillation
cameras for medical applications is currently under study.\end{abstract}
\begin{IEEEkeywords}
position reconstruction, scintillation camera, maximum likelihood,
weighted least squares, dark matter, WIMPs, ZEPLIN-III, liquid xenon,
dual phase detectors.
\end{IEEEkeywords}

\section{Introduction}

\thispagestyle{empty}

\IEEEPARstart {A} {number} of applications require measurement
of the interaction coordinates within a particle detector. In the
low energy region $<$1~MeV, these include medical radionuclide imaging,
gamma-ray astronomy and direct dark matter search experiments. In
the latter instance, which motivated the present work, event localization
\emph{per se} is not relevant for detection of dark matter particles,
but position sensitivity is important for efficient reduction of the
radiation background and correct identification of the candidate events.

ZEPLIN-III is a dual phase (liquid/gas) xenon detector built to identify
and measure galactic dark matter in the form of Weakly Interacting
Massive Particles (WIMPs). It operated at the Boulby mine (UK) between
2006 and 2011. The detector measures both the scintillation light
(S1) and the ionisation charge generated in the liquid by interacting
particles and radiation. The ionisation charge drifts upwards to the
liquid surface by means of a strong electric field and is extracted
into a thin layer of gaseous xenon where it generates UV photons by
electroluminescence (S2). Both the scintillation and electroluminescence
light are measured by a PMT array and the ratio between S1 and S2
allows to discriminate nuclear recoils (expected to be produced by
elastic scatter of WIMPs off xenon nuclei) from the electron recoils
from $\beta$ and $\gamma$\nobreakdash-ray backgrounds. The details
on liquid xenon detector technology as well as on operation of dual
phase detectors can be found in recent review papers \citep{aprile_liquid_2010,Chepel2012}.

The self-shielding property of liquid xenon reduces the rate of background
in the interior of the liquid. Using accurate position reconstruction
to select only events in an inner \textquotedbl{}fiducial\textquotedbl{}
volume therefore improves sensitivity to the WIMP signal. While the
depth of the interaction can be inferred very accurately (few tens
of $\mu$m FWHM) from the electron drift time in the liquid (the delay
between S1 and S2), the position in the horizontal plane has to be
reconstructed from the light distribution pattern across the PMT array.
Another reason for analysis of the light distribution is the need
to eliminate the multiple scatter events that can mimic the WIMP interactions
if one of the scatters has occurred in a dead volume of liquid xenon
from where no charge can be extracted.

The active volume of ZEPLIN-III is a flat layer of liquid xenon (\ensuremath{\approx}40~cm
in diameter and 3.6~cm thick) above a compact hexagonal array of
31 2-inch vacuum ultraviolet-sensitive PMTs (ETL D730/9829Q) immersed
directly in the liquid \citep{Akimov2007}. Such a flat geometry makes
it (from the point of view of position reconstruction) rather similar
to the well-studied scintillation camera, which is widely used in
areas as diverse as medical research and experimental astrophysics
\citep{Joung2000,Cook1985}. The position of an event in a scintillation
camera is traditionally found by the Anger method which consists in
calculating a centroid of the PMT response \citep{anger_scintillation_1958}.

Statistical reconstruction algorithms by maximum likelihood and weighted
least squares methods have gained popularity following the pioneering
work of Gray and Macovski in 1976 \citep{Gray1976}. They offer better
precision along with the possibility of checking if the input data
correspond to a valid event. These methods require knowledge of the
light response functions (LRF) that characterise the response of a
given PMT as a function of position of an isotropic light source inside
the sensitive volume of the detector. Typically, the LRFs are either
measured directly (e.g. by means of a moving collimated radioactive
source) or calculated from the detector geometry, either analytically
or by means of a Monte Carlo simulation.

In the present work, a method of reconstructing LRFs \textit{in situ}
from the calibration data obtained by irradiating the detector by
$\gamma$\nobreakdash-rays from an uncollimated radioactive source
was developed. Based on the set of reconstructed LRFs, the positions
and light yields of scintillation events in the detector can be readily
found using either maximum likelihood or weighted least squares methods.
This procedure was applied to the WIMP-search data taken with ZEPLIN-III
\citep{Lebedenko2009,Lebedenko2009a,Akimov2010,akimov_wimp-nucleon_2012}.

\section{Experimental Setup}

\selectlanguage{english}%
\begin{figure}[!t]
\centering \includegraphics[bb=20bp 120bp 690bp 500bp,clip,width=3.5in]{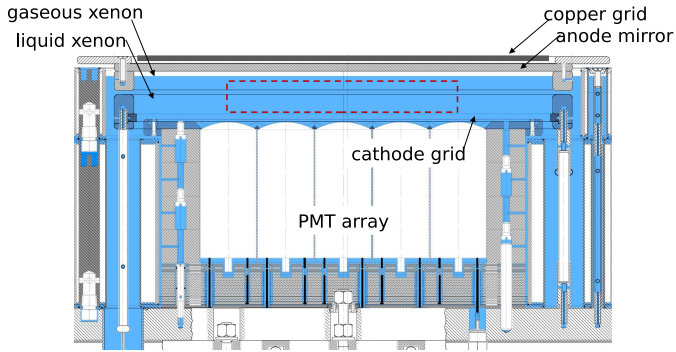}
\caption{\selectlanguage{british}%
\label{fig:Target}Schematic diagram of the ZEPLIN-III WIMP target
region, showing the PMT array and anode and cathode defining the active
volume. Liquid xenon is shown in blue. The dashed box illustrates
the fiducial volume used for WIMP searches.\selectlanguage{english}%
}
\end{figure}

\selectlanguage{british}%
The target region of the detector is shown in Fig. \ref{fig:Target}.
The electric field in the active xenon volume (3.9~kV/cm in the liquid
and 7.8~kV/cm in the gas) is defined by a cathode wire grid 36~mm
below the liquid surface and an anode plate in the gas phase, 4~mm
above the liquid. A second wire grid is located 5~mm below the cathode
grid just above the PMT array. This grid defines a reverse field region
which suppresses the collection of ionisation charge for events just
above the array and helps to isolate the PMT input optics from the
external high electric field. The PMTs are powered by a common high
voltage supply, with the outputs roughly equalised by means of attenuators
(Phillips Scientific 804). The PMT signals are digitised at 2~ns
sampling by 8-bit flash ADC (ACQIRIS DC265). To expand the dynamic
range of the system, each PMT signal is recorded by two separate ADC
channels: one directly and one after amplification by a factor of
10 by fast amplifiers (Phillips Scientific 770). The acquired waveforms
were analysed by a dedicated software that searched for pulses above
a certain threshold and stored them in a parametrised form \citep{Neves2011}.
Subsequently, an event filtering tool was used to retain events with
a fast S1 signal preceding a wider S2 one. All multiple scatter events
containing more than one S2 are filtered out.

\begin{figure}[!t]
\centering \includegraphics[width=3.5in]{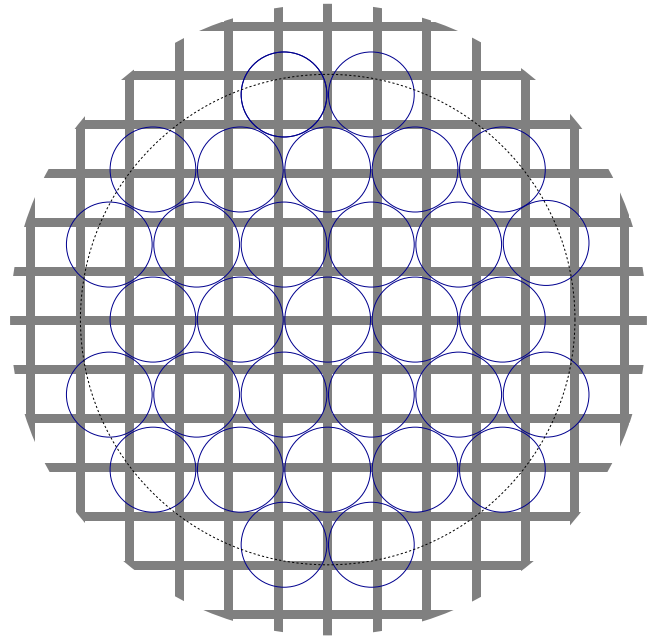}
\caption{\selectlanguage{british}%
\label{fig:The-PMT-array}The PMT array and the copper grid, viewed
from the top. The 31 photomultiplier envelopes are represented by
blue circles. The dashed circle has 150~mm radius.\selectlanguage{english}%
}
\end{figure}

A $^{57}$Co radioactive source was used for calibrating the energy
response of the detector. This source emits 122~keV and 136~keV
$\gamma$\nobreakdash-rays which are rapidly absorbed in liquid xenon
(with attenuation length $<$~4~mm for these energies) mostly by
photoelectric capture \citep{XCOM}. Consequently, most of the interactions
can be considered point-like with full energy deposit. The source
was positioned at approximately 190~mm above the liquid surface and
as close as possible to the detector axis. The calibration was performed
daily to monitor the detector stability. There were also several dedicated
runs aimed at acquiring sufficient data to train the positioning algorithms.
Before the second science run, a specially-designed rectangular copper
grid was placed inside the chamber, above the sensitive volume (Fig.
\ref{fig:The-PMT-array}). The grid structure is 386~mm in diameter,
and was manufactured by diamond wire cutting from a 5.1~mm thick
copper plate; the void pitch is 30~mm and the straight sections are
5~mm wide. The thickness of the grid was chosen such that it would
attenuate the $\gamma$\nobreakdash-ray flux from the calibration
source by approximately a factor of 2, creating a shadow image that
can be used to verify and fine-tune position reconstruction.

\selectlanguage{english}%

\selectlanguage{british}%

\section{Event reconstruction methods}

The problem of event reconstruction consists in finding the energy
(or, rather, the light signal intensity $\hat{N}$) and the position
of an event $(\hat{x},\hat{y})$ given a set of the corresponding
PMT output signals $A_{i}$. For an event at position $\mathbf{r}$
producing $N$ photons the probability of the $i$-th PMT detecting
$n_{i}$ photons is well approximated by the Poisson distribution
\citep{Barrett2009}: 
\begin{equation}
\text{P}_{i}(n_{i})=\frac{\mu_{i}^{n_{i}}e^{-\mu_{i}}}{n_{i}!}\label{eq: Poisson}
\end{equation}
 where $\mu_{i}=N\eta_{i}(\mathbf{r})$ is the expectation for a number
of photons detected by the $i$-th PMT out of $N$ initial ones with
$\eta_{i}(\mathbf{r})$ being LRFs -- the fraction of the photons
emitted by a light source at position $\mathbf{r}$ that produce a
detectable signal in the $i$-th PMT. The corresponding output signal
$A_{i}$ in the general case is a random variable with an expectation
value proportional to $n_{i}$. The probability distribution for $A_{i}$
depends on the single photoelectron response of the corresponding
PMT and can be quite complex \citep{Prescott1966,Gale1966}. However,
in a few special cases it can be approximated by simple functions.
These special cases include:
\begin{itemize}
\item Photon counting. If $n_{i}$ is small (say, less than 10) and the
PMT has a narrow single photoelectron distribution then $n_{i}$ can
be calculated (almost) unambiguously from $A_{i}$  by rounding the
ratio \foreignlanguage{english}{$A_{i}/q_{si}$}, where $q_{si}$
is the average single photoelectron response of the PMT.
\item Normal distribution. If $n_{i}$ is large (say, 25 or more) and the
single photoelectron distribution of the PMT is reasonably symmetric
then, following from the central limit theorem, $A_{i}$ is approximately
normally distributed with the mean equal to $n_{i}q_{si}$.
\end{itemize}

\subsection{Centroid and corrected centroid}

The centroid method of position estimation is the oldest method used
by Anger in the first gamma camera in 1957 \citep{anger_scintillation_1958}.
It is still widely in use due to its simplicity and robustness. The
position estimate is found as the weighted average of PMT coordinates
with weights determined by the light distribution across the PMT array:
\begin{equation}
\hat{x}=\frac{\sum_{i}X_{i}A_{i}f_{i}}{\sum_{i}A_{i}f_{i}}\,,\quad\hat{y}=\frac{\sum_{i}Y_{i}A_{i}f_{i}}{\sum_{i}A_{i}f_{i}}\,,\label{eq:Centroid}
\end{equation}
where $(X_{i},Y_{i})$ are the coordinates of the axis of i-th PMT,
$A_{i}$ is the measured charge and $f_{i}$ is a flat-fielding coefficient
which compensates for variations in gain and quantum efficiency across
the PMT array. As one can see from equations \eqref{eq:Centroid},
no information on LRFs and $A_{i}$ probability distribution is necessary
for application of this method. On the other hand, while the centroid
method works reasonably well close to the centre of the detector (up
to 100 mm from the centre in ZEPLIN-III), it becomes increasingly
biased for events in the periphery. Another disadvantage is that it
gives no indication regarding the match of the actual light distribution
to the expected one.

If there exists one-to-one mapping between the true position and the
one reconstructed by the centroid method then it is possible to invert
this mapping to obtain the unbiased \textquotedbl{}corrected\textquotedbl{}
estimate from the biased centroid one. In practice, this is often
done by building a look-up table for a number of known positions on
a rectangular grid and interpolating between these points. Another
possibility is to use Monte Carlo simulation to calculate the forward
mapping and then to use numerical methods to invert it. The latter
method was employed in the ZEPLIN-III event filtering routine. It
was also used to obtain the first approximation in the iterative LRF
reconstruction procedure.

\subsection{Maximum likelihood}

The maximum likelihood (ML) technique \citep{Joung2000,Cook1985,Clinthorne1987}
consists in finding the set of parameters that maximises the likelihood
of obtaining the experimentally measured outcome. For the case of
photon counting when $n_{i}$ are known for each PMT, the likelihood
function can be easily calculated from the Poisson distribution \eqref{eq: Poisson}:
\begin{equation}
\ln L=\sum_{i}\ln P(n_{i},\mu_{i})=\sum_{i}(n_{i}\ln\mu_{i}-\mu_{i})-\sum_{i}\ln(n_{i}!)\,.\label{eq:ML-pois}
\end{equation}

Taking into account that $\mu_{i}=N\eta_{i}(\mathbf{r})$, one can
write \citep{Cook1985}

\begin{equation}
\ln L(\mathbf{r},N)=\sum_{i}\left(n_{i}\ln(N\eta_{i}(\mathbf{r}))-N\eta_{i}(\mathbf{r})\right)+C\,,\label{eq:ML-pois-LRF}
\end{equation}
where $C$ does not depend on neither $\mathbf{r}$ or $N$. If the
LRFs $\eta_{i}(\mathbf{r})$ are known, the best estimates $\mathbf{\hat{r}}$
and $\hat{N}$ can be found in a straightforward way by maximising
function \eqref{eq:ML-pois-LRF}. The best estimate of $N$ at given
$\mathbf{r}$, $\hat{N}(\mathbf{r})$ can be found analytically:

\begin{equation}
\hat{N}(\mathbf{r})=\frac{\sum_{i}n_{i}}{\sum_{i}\eta_{i}(\mathbf{r})}\,.\label{eq:ML-pois-N-estimate}
\end{equation}
By substituting $\hat{N}$ for $N$ into \eqref{eq:ML-pois-LRF} one
obtains $\ln L_{m}(\mathbf{r})=\ln L(\mathbf{r},\hat{N}(\mathbf{r}))$,
which is a function of the position only. Then $\hat{N}$ and $\mathbf{\hat{r}}$
are found by maximising $\ln L_{m}(\mathbf{r})$ either analytically
or by numerical methods. As a bonus, for the 2D case $\ln L_{m}(\mathbf{r})$
can be visualised as a colour map, which is very useful for either
debugging or checking the validity of a given event.

\subsection{Weighted least squares}

If $A_{i}$ can be considered normally distributed the more flexible
weighted least squares (WLS) method can be used instead of ML \citep{Ling2008}.
In this case the parameter estimates are found by minimising the weighted
sum of squared residuals $\chi^{2}$:
\begin{equation}
\chi^{2}=\sum_{i}w_{i}(A_{ei}-A_{i})^{2}\,,\label{eq:WLS-0}
\end{equation}
where $A_{ei}=\mu_{i}q_{si}=N\eta_{i}(\mathbf{r})q_{si}$ is the expected
PMT output charge and $w_{i}$ is the weighting factor which is reciprocal
to the variance of $A_{ei}-A_{i}$. The best estimates $\mathbf{\hat{r}}$
and $\hat{N}$ are obtained by finding the global minimum of
\begin{equation}
\chi^{2}(\mathbf{r},N)=\sum_{i}w_{i}(\mathbf{r},N)\left(N\eta_{i}(\mathbf{r})q_{si}-A_{i}\right)^{2}\,.\label{eq:WLS}
\end{equation}
The $N$ and $\mathbf{r}$ minimisations can be separated, as in the
likelihood case, reducing by one the dimensionality of the problem. 

Under an assumption that $A_{i}$ is measured exactly and the variance
of $A_{ei}$ is only due to statistical fluctuations in the number
of detected photoelectrons, the WLS method becomes equivalent to ML
\citep{cowan_statistical_1998}. However, in a real detector the measured
$A_{i}$ differs from the true value because of electronic noise.
The variance of $A_{ei}$ is also typically higher then expected from
Poisson statistics due to non-zero width of the single photoelectron
distribution. Compared to the ML method, the WLS makes it much easier
to account for these and other factors. Most importantly, it makes
it possible to reduce the weights for those PMTs with less well known
light response.

\begin{figure*}
\begin{tabular}{rrr}
 \includegraphics[bb=10bp 20bp 350bp 350bp,clip,scale=0.44]{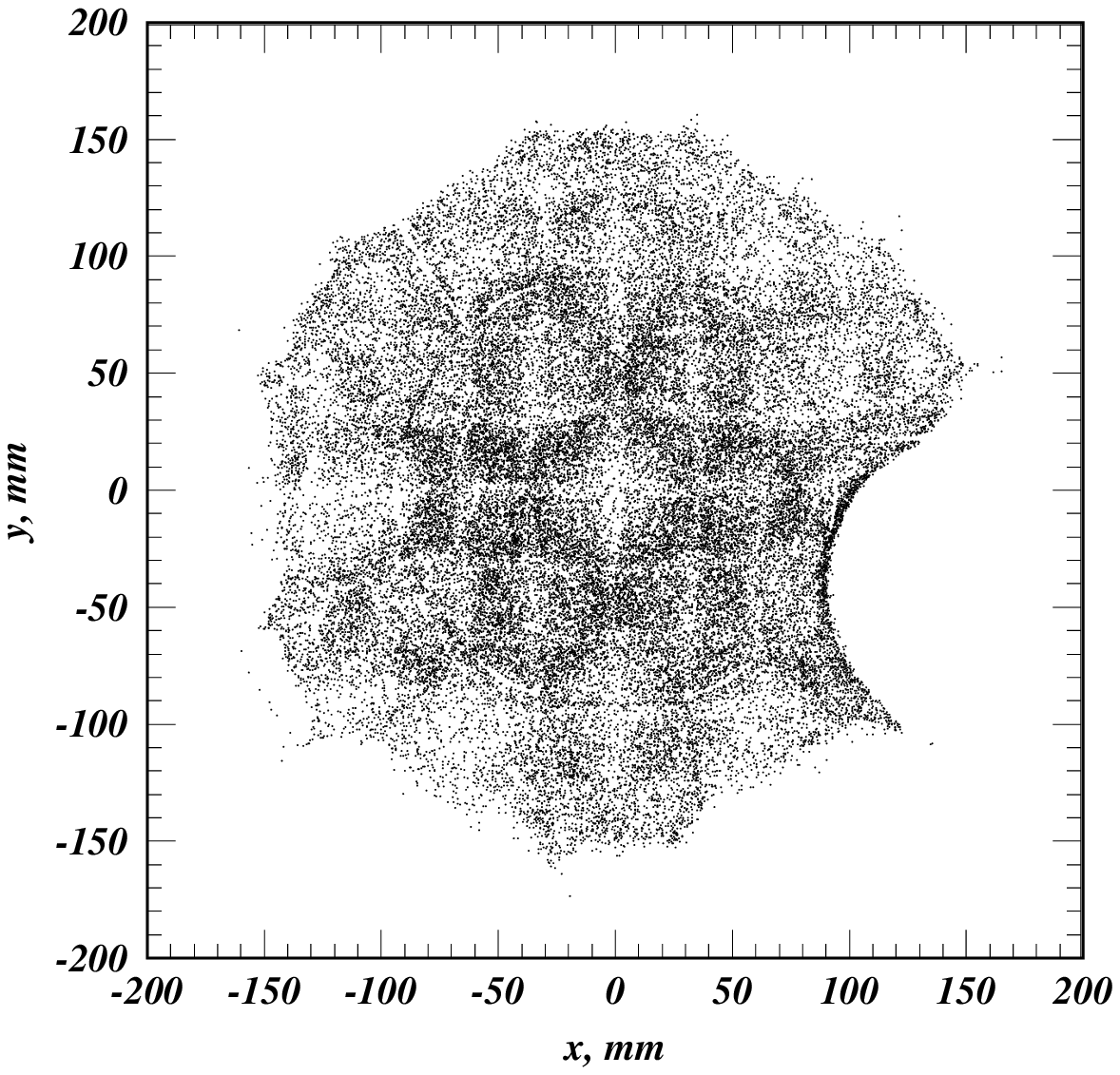} & \includegraphics[bb=10bp 20bp 350bp 350bp,clip,scale=0.44]{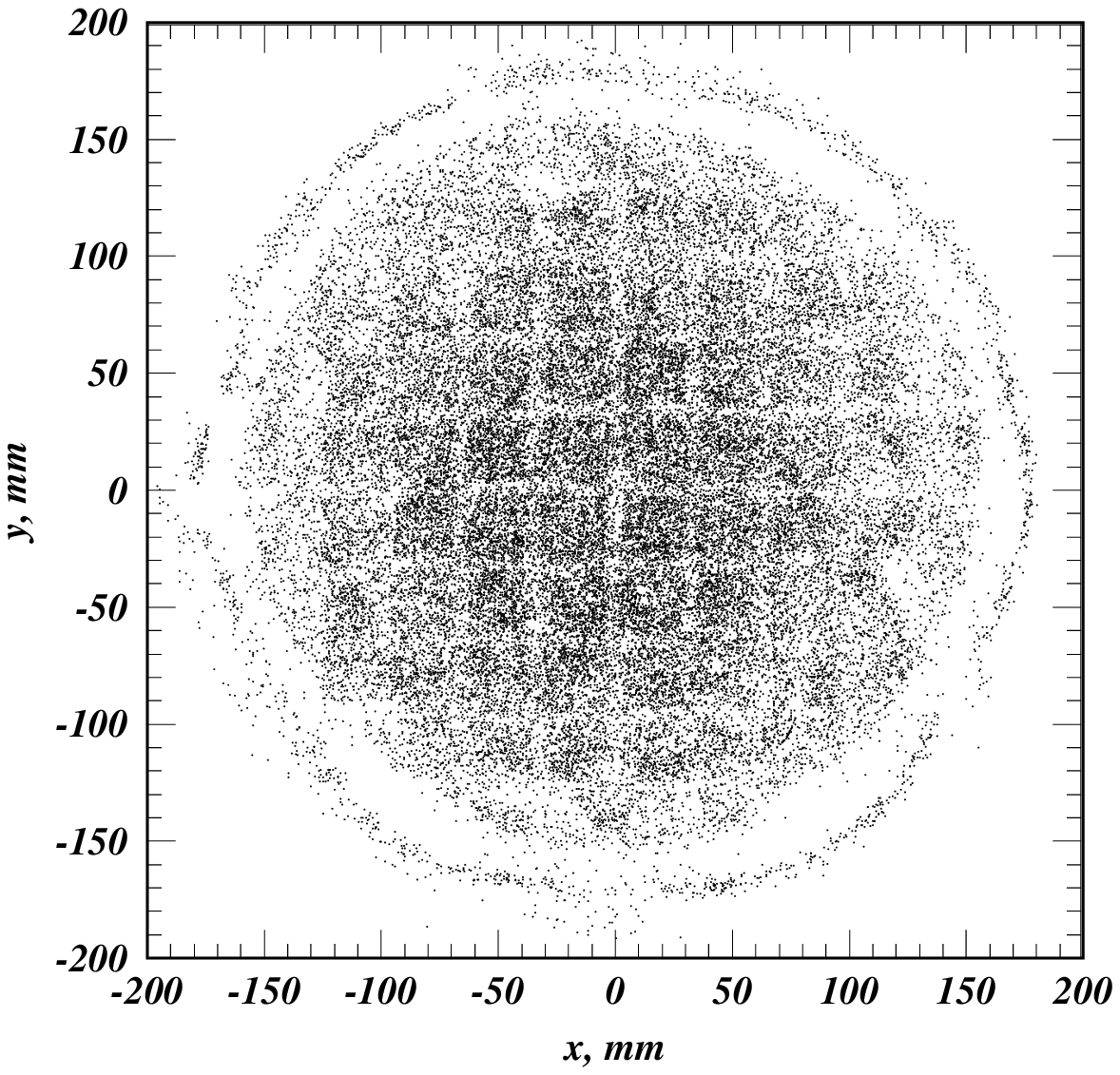} & \includegraphics[bb=10bp 20bp 350bp 350bp,clip,scale=0.44]{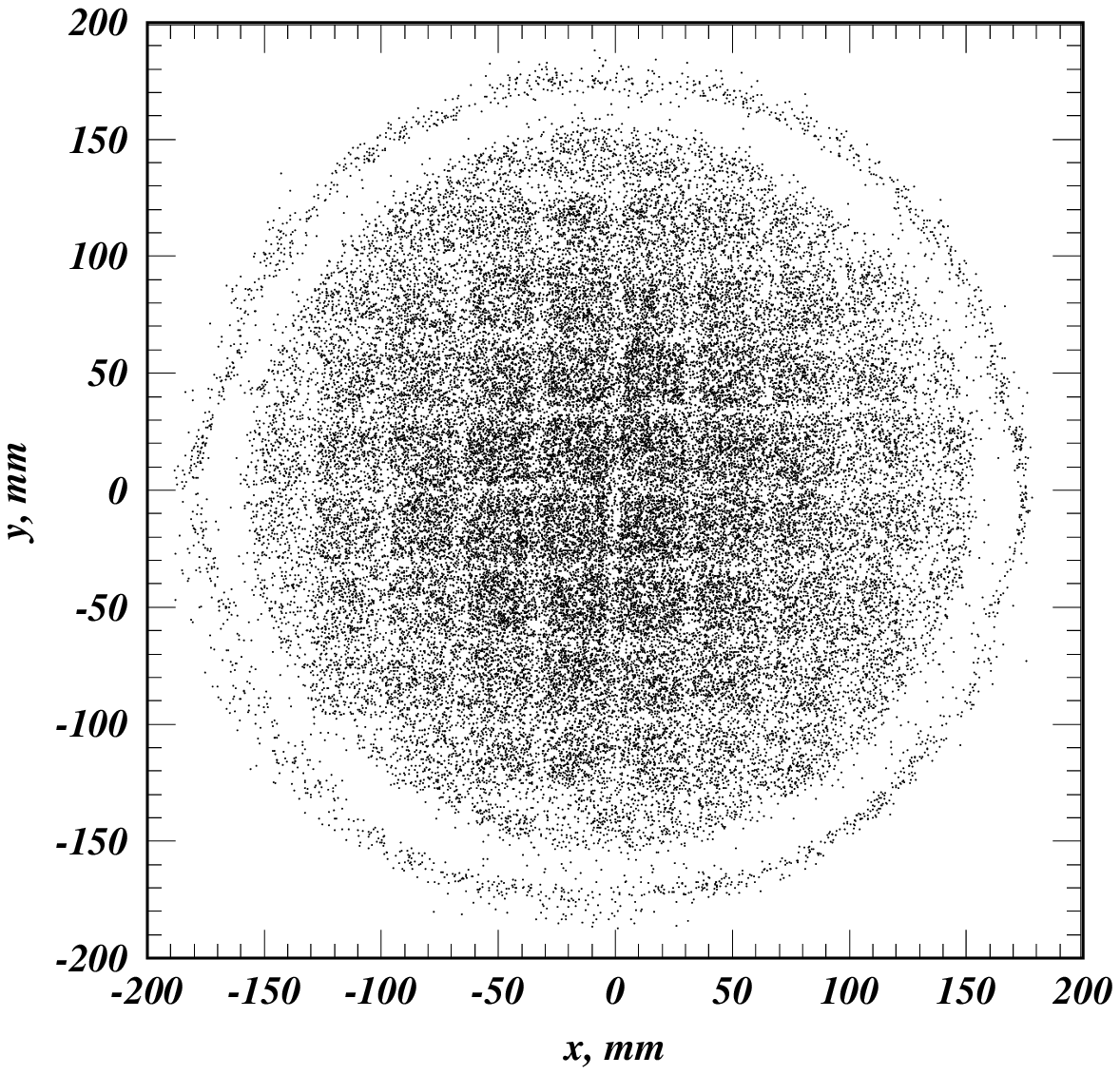}\tabularnewline
a)\includegraphics[bb=10bp 20bp 350bp 350bp,clip,scale=0.44]{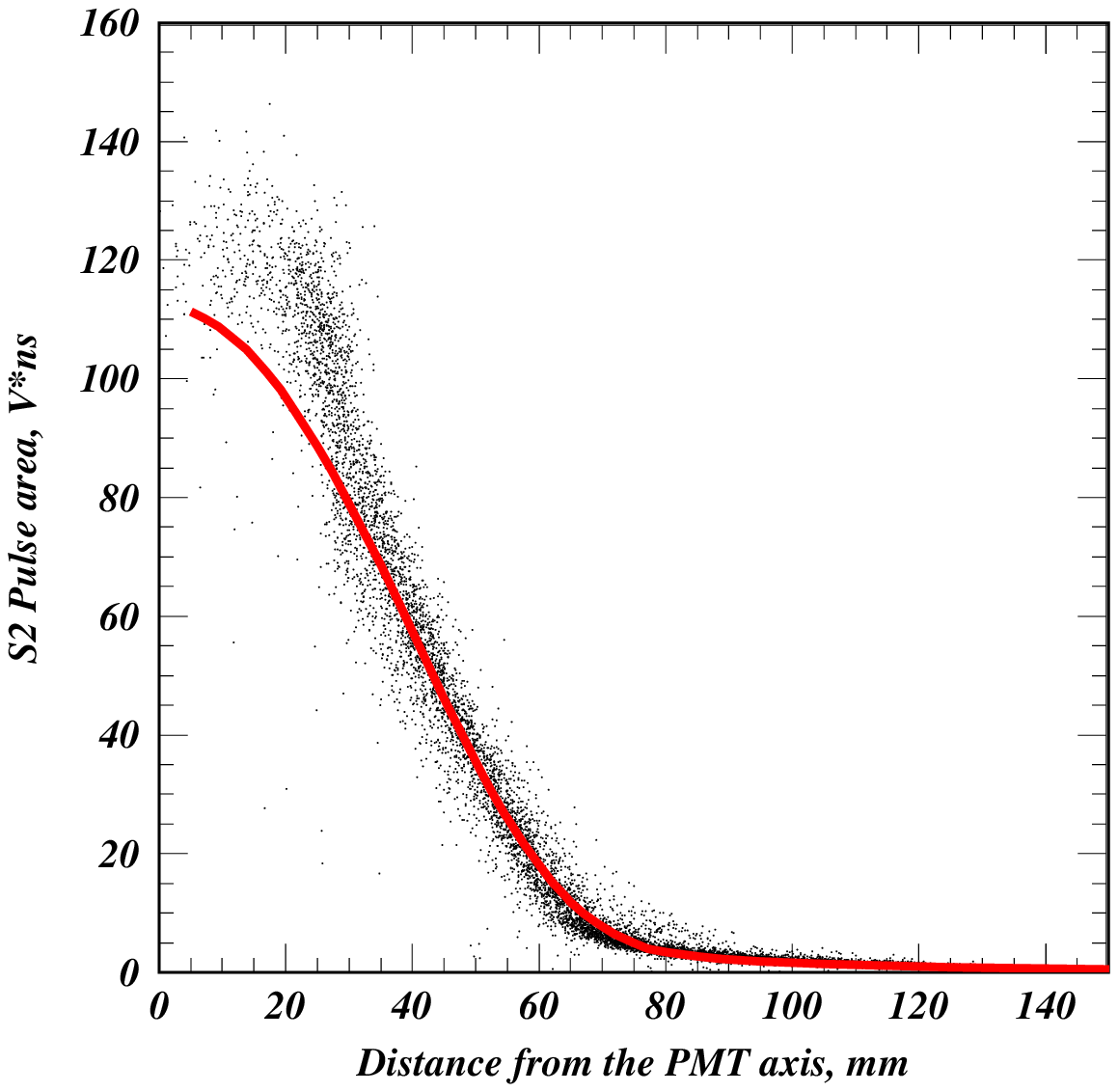} & b)\includegraphics[bb=10bp 20bp 350bp 350bp,clip,scale=0.44]{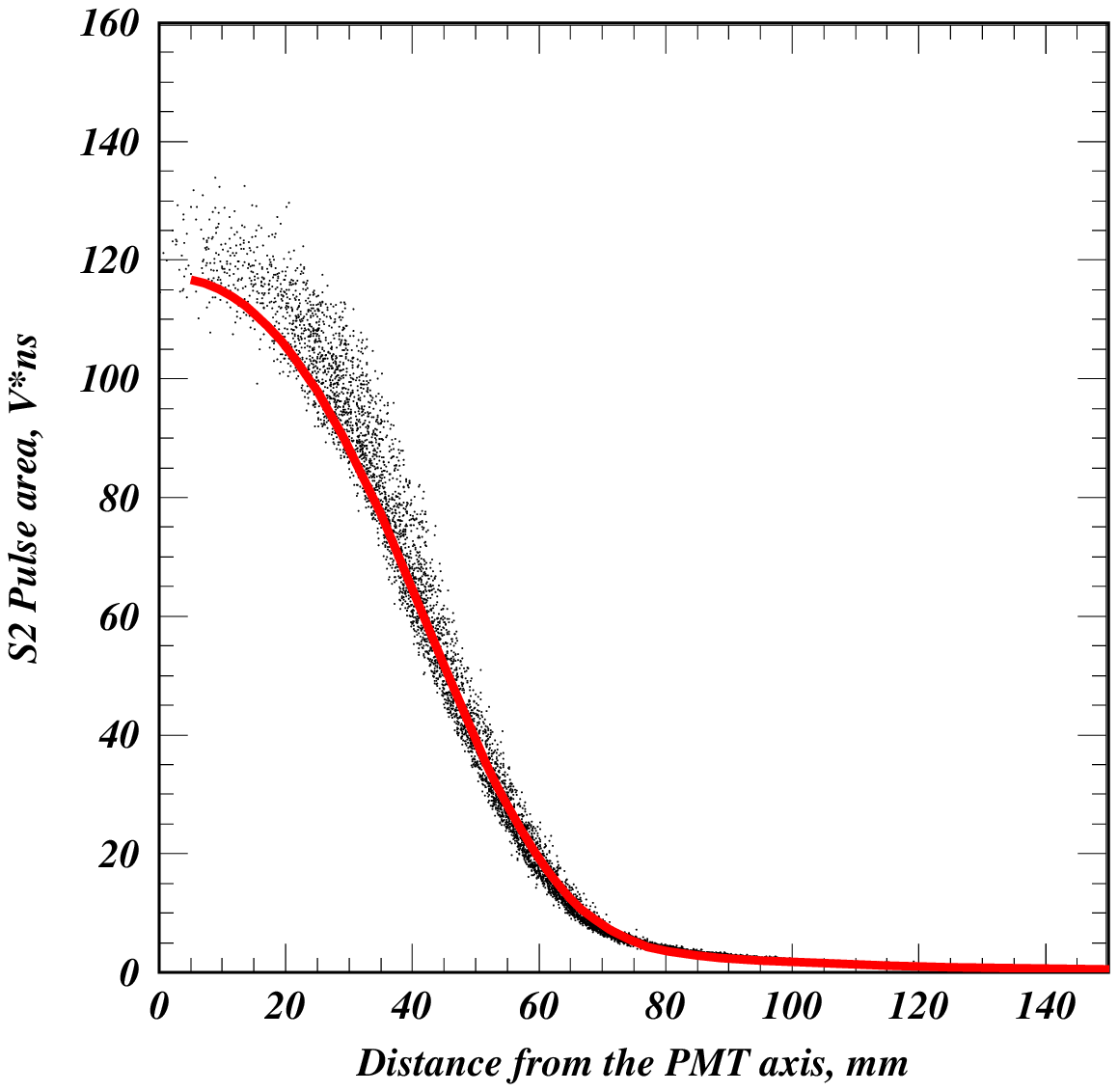} & c)\includegraphics[bb=10bp 20bp 350bp 350bp,clip,scale=0.44]{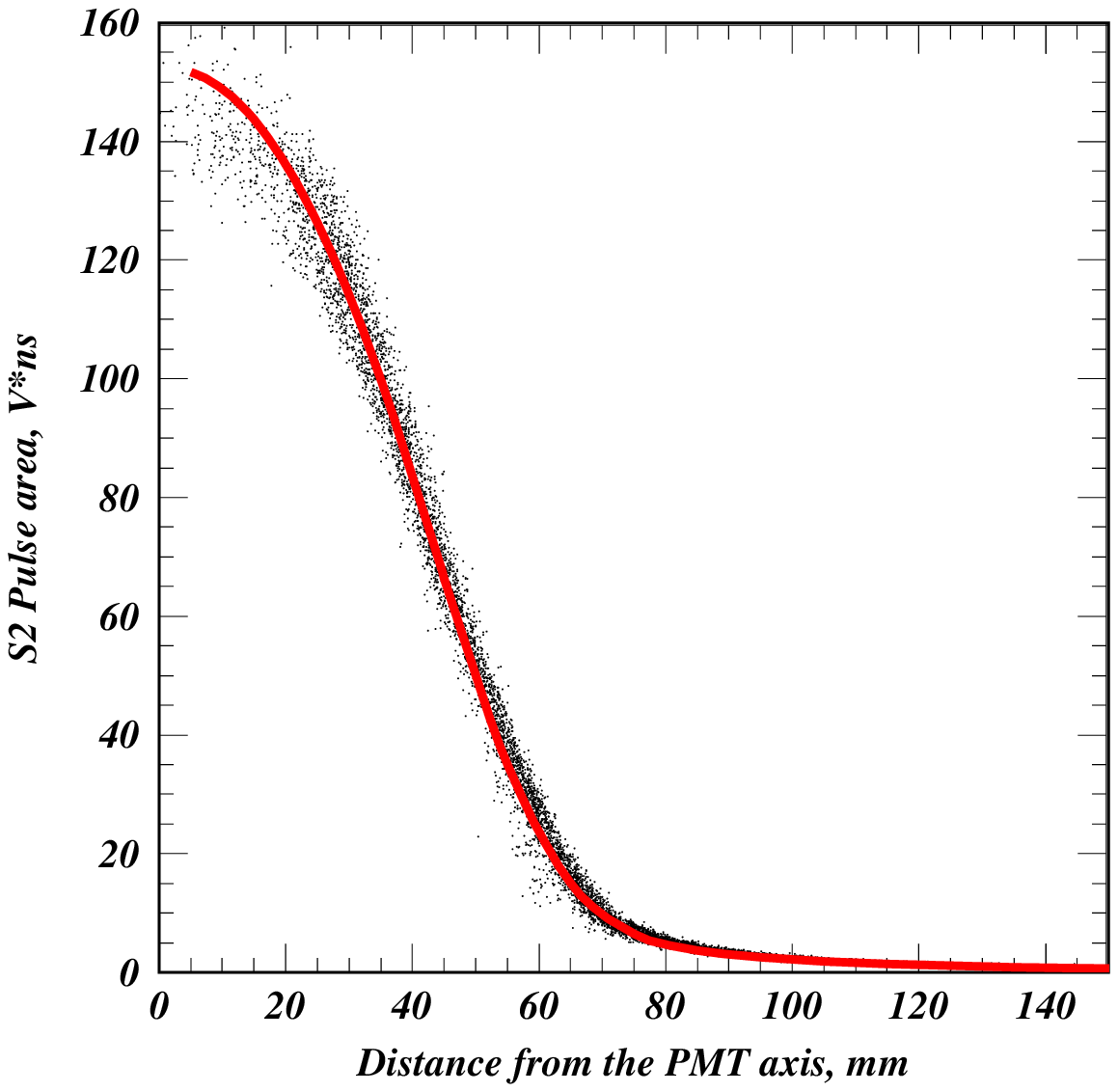}\tabularnewline
\end{tabular}

\caption{\label{fig:LRF-Reconstruction}Iterative reconstruction of the LRFs
from $^{57}$Co calibration data. The top row: the evolution of the
distribution of estimated event positions from S2 pulses. The bottom
row: the response of PMT~11 (with centre at ($-$79.5, $-$45.9))
versus estimated distance from its centre (dots) and the corresponding
S2 LRFs derived from these distributions (curve). a) Initial position
estimates obtained by centroid. b) First iteration. c) Final (5\protect\nobreakdash-th)
iteration.}
\end{figure*}

\subsection{Method choice}

The choice of the WLS method for S2 reconstruction is straightforward:
due to its high light output, the S2 signal statistic is quasi-normal
except for the PMTs far from low energy events. These PMTs may be
either ignored or clustered together so that the photoelectron statistic
per cluster is quasi-normal too. 

In the case of S1, the total collected charge (from the whole PMT
array) is equivalent, depending on the event position, to 1--2 photoelectrons
per keV; this means that in the region of interest for WIMP searches
(<50~keV) the S1 distribution is too far from normal to use the WLS
method with confidence. Consequently, the ML method was used.

\section{Reconstruction of light response functions}

The ML and WLS methods described above rely on t\textcolor{black}{he
knowledge of} the LRFs $\eta_{i}(\mathbf{r})$. There are several
methods for obtaining the LRFs described in the literature. The most
straightforward of these is the direct measurement, by scanning the
detector with a moving well-collimated $\gamma$-ray source \citep{Barrett2009,milster_digital_1985}.
Unfortunately, a combination of several factors made this method \textcolor{black}{impractical}
for the ZEPLIN-III detector. Because of the cryostat, the source could
not be placed closer than 190 mm to the liquid surface, and therefore
a long collimator was required to achieve good position resolution.
However, the available space above the detector was extremely limited
and installation of a scanning system would require reduction in the
shield thickness that was deemed unacceptable. 

Alternatively, the LRFs can be calculated from the detector geometry
by means of a Monte Carlo simulation tuned to reproduce the experimental
data \citep{neves_position_2007}. Although we attempted a similar
method with ZEPLIN-III \citep{lindote_preliminary_2007}, the difficulty
in reproducing the exact shape of the LRFs by Monte Carlo and time
constraints limited its use.

Yet another approach, explored in \citep{Ling2008}, is to choose
a suitable parametric model for the LRF and adjust the parameters
so as to minimize the mean sum of squared residuals in the WLS method
for a population of calibration events. This method works well for
low parameter count models proposed in that study. However, the LRF
shape for S2 in the ZEPLIN-III detector proved to be more complex,
requiring a model with at least five parameters in the simplest case:
\begin{equation}
\eta(\rho)=A\exp\left(-\frac{a\rho}{1+\rho^{1-\alpha}}-\frac{b}{1+\rho^{-\alpha}}\right),\;\rho=r/r_{0}\,,\label{eq:Z3-param-model}
\end{equation}
where $r$ is the distance from the PMT axis and $A$, $r_{0}$, $a$,
$b$ and $\alpha$ are adjustable parameters. As a result, the parameter
adjustment becomes much more difficult due to increased dimensionality
of the problem. To overcome this difficulty we developed an iterative
method of LRF reconstruction described below.

\subsection{Method description}

In this method the detector is irradiated by a non-collimated monoenergetic
gamma source and the PMT responses are recorded event by event. Even
if the gamma source is not collimated, it is still possible to obtain
an estimate for each event position using the centroid or the corrected
centroid method, at least for the central part of the detector. After
a sufficiently large event sample is acquired, making an additional
assumption that the LRF depends smoothly on $\mathbf{r}$ and assuming
that all the events produce the same amount of light, one can obtain
the first approximation for the LRF $\eta_{i}^{(1)}(\mathbf{r})$
by fitting the PMT response to the events at different $\mathbf{r}$
by a smooth function of $\mathbf{r}$. 

This first approximation can now be used to obtain better estimates
for the positions of the events in the sample using ML or WLS method.
Compared to the centroid estimates, these new estimates are less biased,
especially in the case of peripheral events. Fitting again the PMT
response as a function of coordinates using the updated event positions
gives a second approximation $\eta_{i}^{(2)}(\mathbf{r})$.

The above steps are repeated until some convergence criterion is reached.
This can be the fact that the reconstructed dataset has attained some
quality that the physical calibration events are known to possess,
for example monoenergeticity or some known distribution in the \textit{$xy$}
plane. Another option is to iterate until the change in the LRFs on
the next step falls below a pre-defined tolerance.

Some additional regularization may be necessary to force the iteration
to converge. One is the choice of a smoothing function. Another is
the use of some \emph{a priori} known property of the LRF; for example
in the case of a PMT with a circular photocathode it is reasonable
to assume that the LRF has axial symmetry $\eta(\mathbf{r})=\eta(r)$,
where $r$ is the distance from the PMT axis. This type of regularization
was used in LRF reconstruction for ZEPLIN-III, the applicability of
it will be discussed in section \ref{sub:Discussion}.

\subsection{ZEPLIN-III example}

In order to collect the data necessary for reconstruction of the S2
LRFs, the detector was irradiated with $\gamma$\nobreakdash-rays
from a $^{57}$Co source. The top plot in Fig.~\ref{fig:LRF-Reconstruction}(a)
shows the $x$-$y$ distribution of the estimated $^{57}$Co event
positions obtained with a corrected centroid algorithm. Clearly, the
events on the periphery tend to be misplaced closer to the centre
of the PMT array. The situation deteriorates in the bottom-right corner
where one of the PMTs was not functioning. However, for the central
part of the array, approximately up to 100~mm from the centre, the
centroid performance is good enough to be used for reconstructing
the first approximation for the LRFs. This is demonstrated in the
bottom plot of Fig.~\ref{fig:LRF-Reconstruction}(a), where the area
of PMT response is plotted versus the distance from its axis, calculated
from the event position estimated by centroid. The resulting scatter
plot was fitted using linear least squares technique with a cubic
spline (the smooth curve on the plot) which was used as a first approximation
$\eta_{i}^{(1)}(r)$ for the LRF for a given PMT. Then the set of
LRFs obtained in this way was used to re-calculate positions of the
$\gamma$-ray interactions using the WLS method, producing the position
distribution shown on the top plot of Fig.~\ref{fig:LRF-Reconstruction}(b),
and the cycle was repeated. After 5 iterations, the LRFs converged
to the final shape shown in Fig.~\ref{fig:LRF-Reconstruction}(c). 

As one can see, the final distribution of the estimated event positions
clearly shows the projected image of the copper grid with no significant
distortions even in the region close to the non-functioning PMT. Note
the ring of events in the periphery of the detector. The analysis
indicates that they are well reconstructed as the sum of squared residuals
$\chi_{min}^{2}$ is compatible with that for the events from the
main population. Our interpretation is that these events occured near
the edge of the field cage where non-uniform electric field pushed
the extracted charge even further to the periphery. While the centroid
algorithm fails to separate them from the main population (they are
actually reconstructed closer to the center than some other peripheral
events), the WLS method allows to unambiguously identify them.

\subsection{\label{sub:Discussion}Discussion}

The important advantage of the method described above is its ability
to handle many more parameters than it was required by the original
five-parameter model \eqref{eq:Z3-param-model}. This means that one
can use (together with appropriate regularization) much more flexible
non-parametric LRF representations such as look-up table \citep{milster_digital_1985}
or cubic spline \citep{joung_cmice:_2002} previously used only in
conjunction with the direct scan method. The cubic spline has an additional
advantage of being a smooth function and, as we have found, with appropriate
choice of knots it does not require any additional regularization.
For this reason, a cubic spline representation for axially symmetric
LRFs $\eta(r)$ was adopted for both S1 and S2 LRFs. The knot placement
was adjusted experimentally to cover the region of most rapid change
in the response function with a denser grid. 

Naturally, the assumption about axial symmetry of the PMT response
is only an approximation. Several factors, most notably non-uniformities
of the PMT photocathode and spatial dependence of the light collection
efficiency (especially near the detector edge) can produce considerable
deviations from symmetry. Such deviations from the model lead to systematic
errors in estimated event position and energy. Fortunately, for such
poorly reconstructed events the minimized sum of squared residuals
$\chi_{min}^{2}$ tends to be above average. Plotting $\chi_{min}^{2}$
against $x$ and $y$ for a calibration dataset reveals the areas
of the detector where the actual light response does not conform to
the model (or rather the model is not good enough). Examining such
plots for ZEPLIN-III we have found no increase in $\chi_{min}^{2}$
value near the detector edge which means that the light collection
efficiency is indeed axially symmetric for both inner and outer PMTs.
This is explained by poor reflectivity of the detector construction
materials to xenon scintillation light ($\lambda\approx$175nm) and
confirmed by Monte Carlo simulations. 

On the other hand, LRF of several PMTs have shown deviation from axial
symmetry at small $r$, most probably due to photocathode non-uniformity.
This was mitigated by introducing uncertainty $\delta_{i}(r)$ for
the corresponding LRF $\eta_{i}(r)$ that effectively reduced the
weight function $w_{i}$ in the sum \eqref{eq:WLS} (in other words,
the contribution of i-th PMT) in the regions where its response was
less symmetric. To improve convergence of first iterations, such ``bad''
PMTs can be temporary ignored by setting $w_{i}=0$.

The method described above assumes that every accepted calibration
event produces the same amount of scintillation light, independently
of its position in the detector. In fact, the number of scintillation
photons per event is affected by systematic and statistic fluctuations.
In the case of a well-designed dual phase detector, the systematic
fluctuations are negligible for the following reasons: 
\begin{itemize}
\item the liquid scintillator stays uniform due to convection flow and diffusion;
\item the light yield for S2 depends on the field strength, the gas pressure
(both uniform across the detector sensitive volume) and the gas gap
width, very well controlled by measuring duration of S2 pulses;
\item almost all small-angle scatters can be eliminated by applying a cut
on the width of S2 pulses.
\end{itemize}
As for systematic fluctuations, it was empirically found that those
with up to \textasciitilde{}20\% rms do not prevent correct reconstruction
of the LRFs.

\section{Results}

\subsection{Spatial resolution}

\begin{figure}
\includegraphics[bb=10bp 20bp 350bp 350bp,clip,scale=0.7]{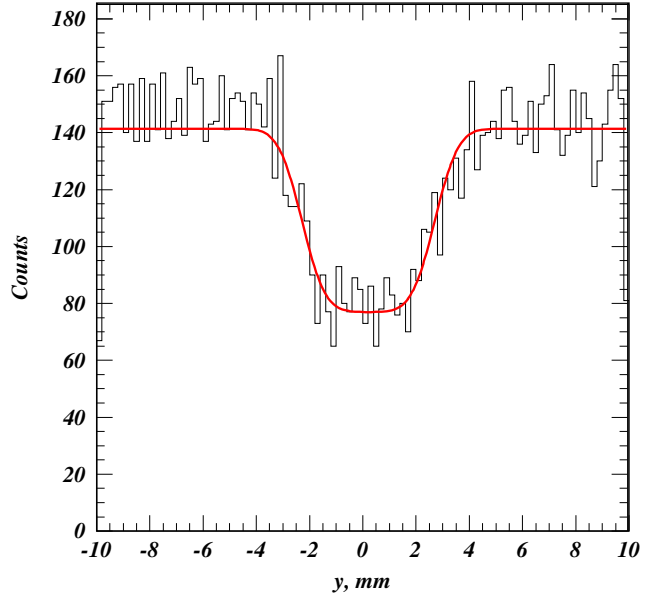}

\caption{\label{fig:Grid-projection}Projection (``shadow'') of the middle
bar of the copper grid, used to estimate the spatial resolution for
S2 for the central part of the detector.}
\end{figure}

\begin{figure}
a)\includegraphics[bb=10bp 20bp 350bp 350bp,clip,scale=0.7]{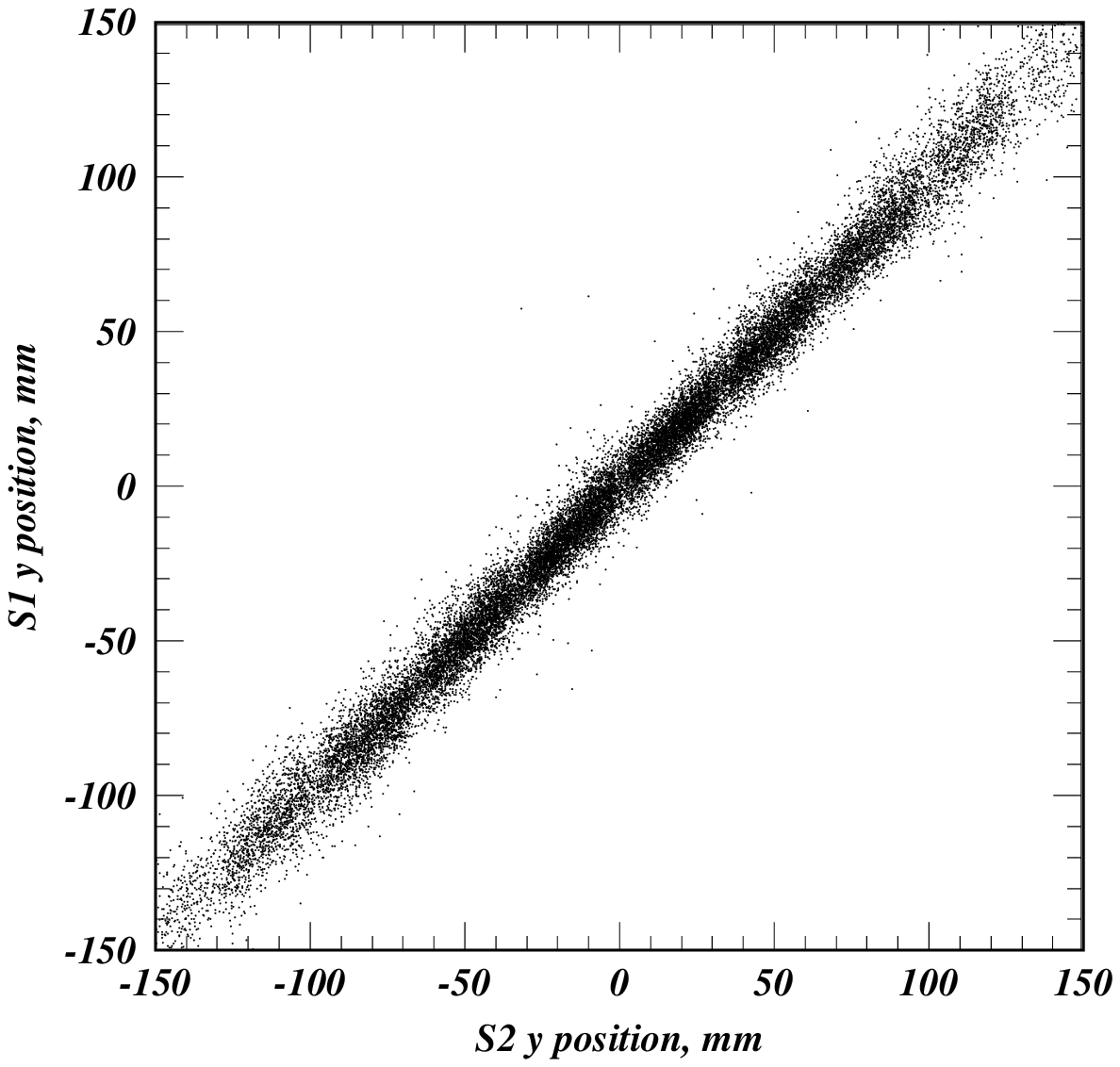}
b)\includegraphics[bb=10bp 20bp 350bp 350bp,clip,scale=0.7]{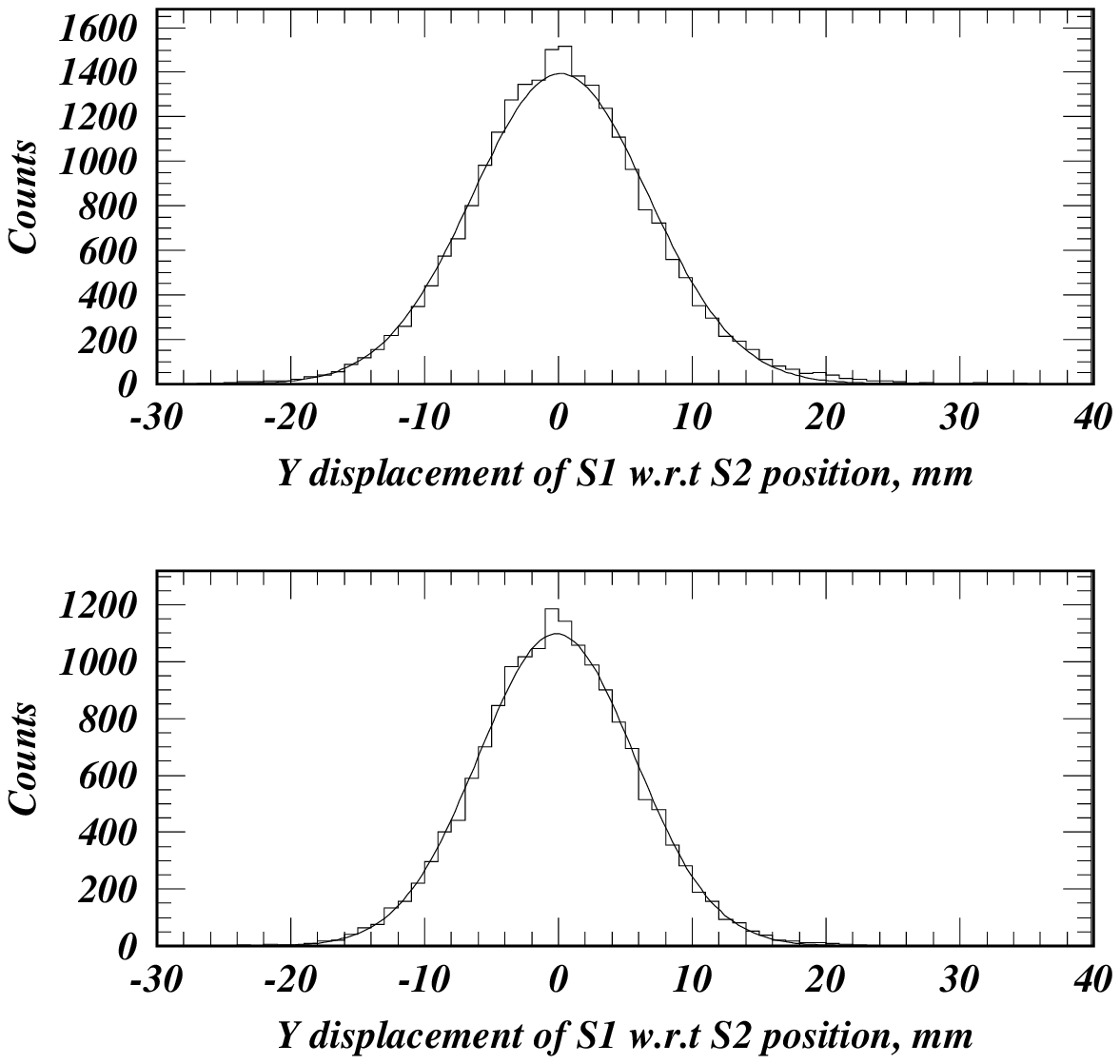}

\caption{\label{fig:S1-spatial-res}(a) The independently reconstructed $y$-coordinates
for S1 and S2 demonstrate, as expected, very strong correlation, (b)
the S1 spatial resolution for the whole fiducial volume (top) and
for the events with $R$$<$100~mm where $R$ is the distance from
the axis of the chamber.}
\end{figure}

Spatial resolution for both S1 and S2 was measured with $^{57}$Co
calibration source. In the central part of the chamber, right below
the source, the $\gamma$\nobreakdash-rays cross the copper grid
at normal incidence creating the sharpest contrast between open and
shadow areas. In the reconstructed event distribution, this transition
is smeared due to finite spatial resolution and, to some extent, by
scattering in the 7-mm anode plate located below the grid. In other
words, the sharpness of the edges of the projected image gives an
upper limit for the spatial resolution of the detector for S2 signals.
In Fig.~\ref{fig:Grid-projection}, the distribution of the $y$-positions
of the reconstructed events is demonstrated for a narrow patch in
the inner part of the detector ($R$$<$100~mm). The distribution
is fitted with a convolution of a step function with the Gaussian
giving resolution of 1.6~mm FWHM. The resolution worsens towards
the edge of the fiducial volume due to combination of lower light
collection and edge effects, becoming \textasciitilde{}3~mm FWHM
at $R$=150~mm.

The spatial resolution for S1 can be estimated by comparing independently
reconstructed coordinates for S1 and S2, Fig.~\ref{fig:S1-spatial-res}(a).
The difference between the two, shown in Fig.~\ref{fig:S1-spatial-res}(b),
is approximately normally distributed with FWHM of 15.0~mm for the
whole fiducial volume and 13.0~mm for events with $R$$<$100~mm.
As the contribution of S2 resolution is obviously negligible, these
values correspond to the spatial resolution for S1. Note that no
energy selection was performed in these measurements so the \textasciitilde{}10\%
admixture of 136 keV present in $^{57}$Co spectrum might marginally
improve the results compared to what would be obtained using a pure
122 keV source.

\subsection{Energy resolution}

\begin{figure}
\includegraphics[bb=10bp 20bp 350bp 350bp,clip,scale=0.7]{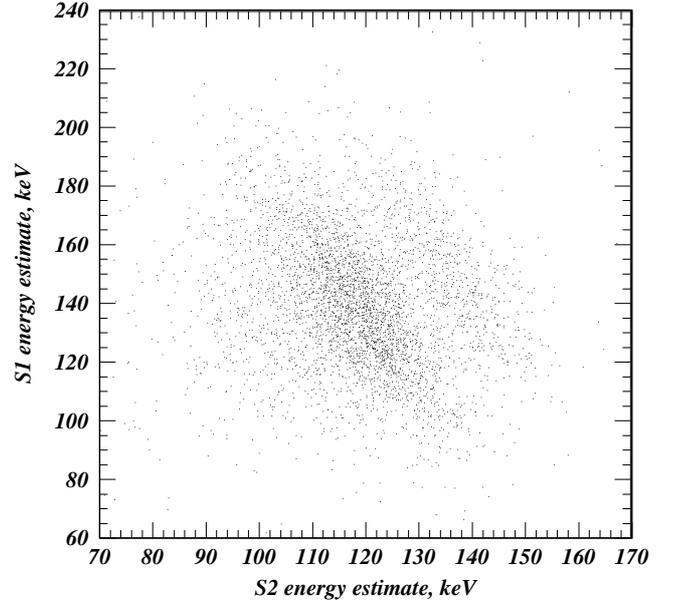}

\caption{\label{fig:Anticorrelation}Anti-correlation between S1 and S2 signals.
Left and right stripes correspond to 122 keV and 136 keV $\gamma$\protect\nobreakdash-rays,
respectively.}

\end{figure}

As demonstrated in \citep{Conti2003}, there is strong anti-correlation
between scintillation light and extracted charge for electron recoils
in liquid xenon under an applied electric field. The reason for this
is that part of the scintillation light comes from recombination.
For the less dense electron tracks, the electron extraction efficiency
is higher while recombination (and scintillation output) is lower.
Thus, fluctuation of the electron track density from event to event
leads to variations in light and charge outputs, which in a dual phase
detector leads in turn to anti-correlated variations of S1 and S2
even for events of the same energy. Consequently, the best energy
estimate for a dual phase detector is a linear combination of S1 and
S2 light outputs. Fig.~\ref{fig:Anticorrelation} shows the relationship
between scaled light outputs for S1 and S2 for the events produced
by $\gamma$\nobreakdash-rays from the $^{57}$Co source. The scaling
factors were chosen so that the mean of the distribution is at 125
units for both S1 and S2. One can see that there is indeed anti-correlation
with S1 varying approximately by a factor of 3 more than S2.

\begin{figure}
a)\includegraphics[bb=10bp 20bp 350bp 350bp,clip,scale=0.7]{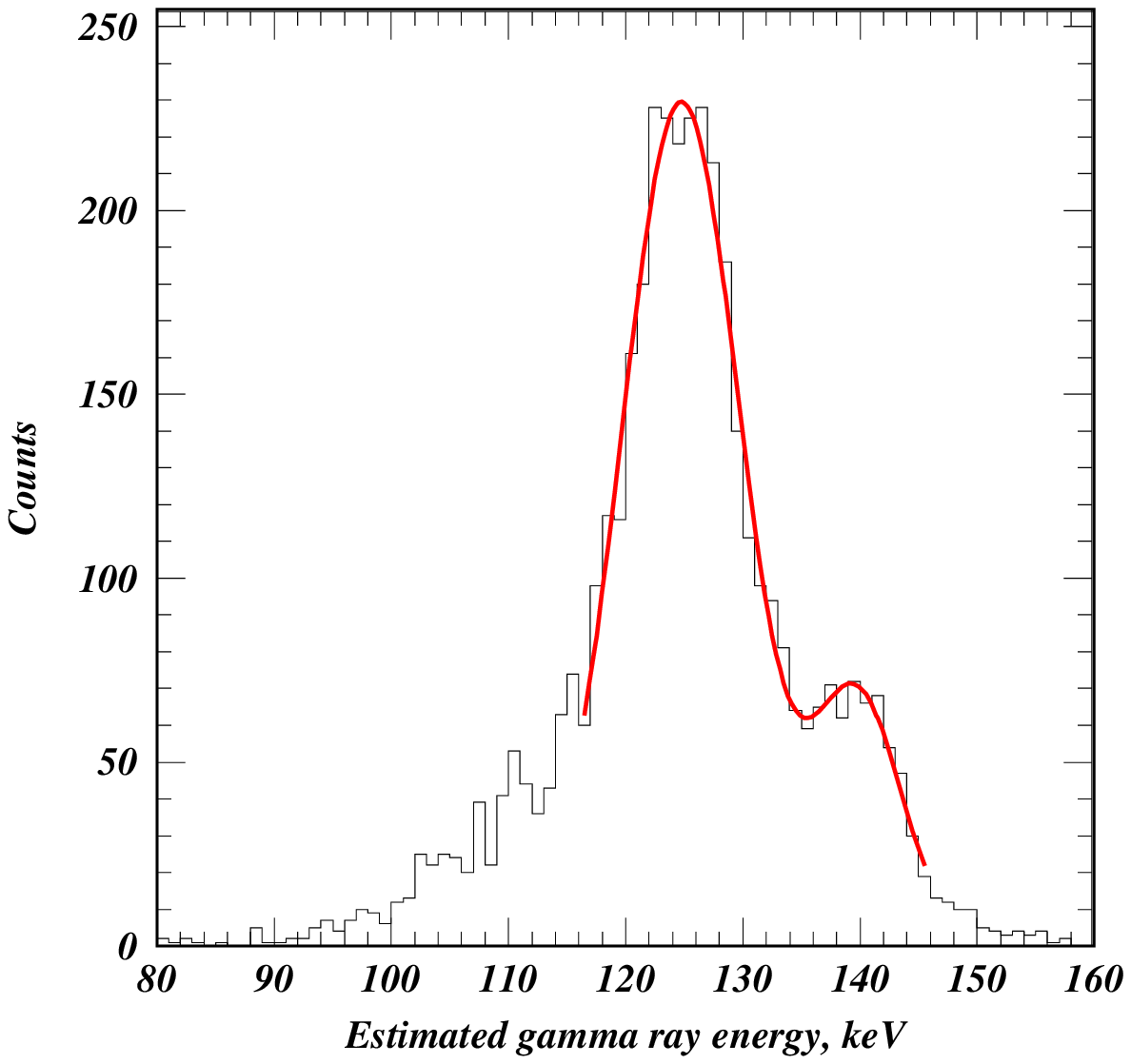}
b)\includegraphics[bb=10bp 20bp 350bp 350bp,clip,scale=0.7]{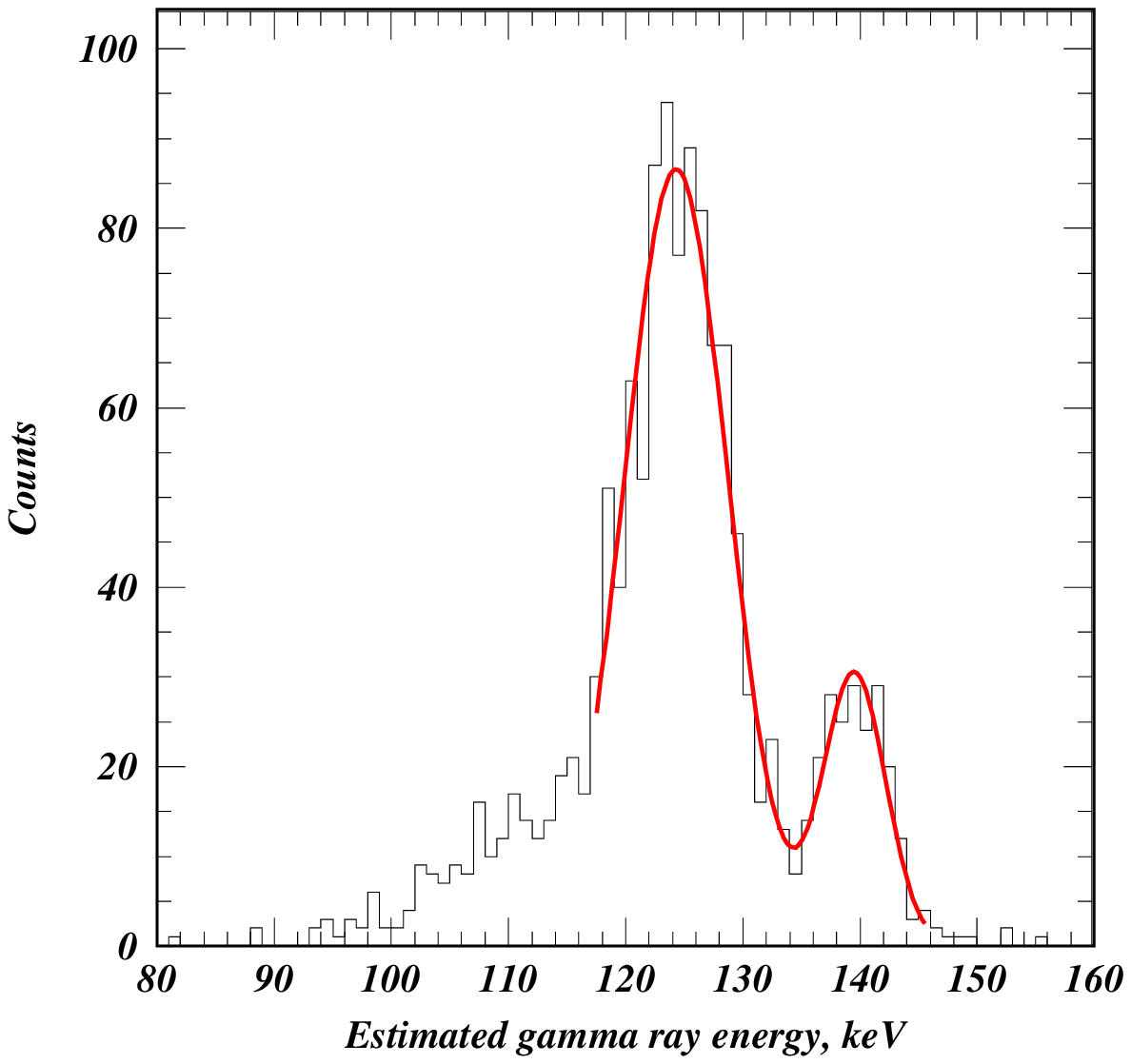}

\caption{\label{fig:EnergyRes}The spectrum of $^{57}$Co $\gamma$-ray energy
estimated from a linear combination of S1 and S2 light yields for
the whole fiducial volume (a) and for the central spot with $R$$<$50~mm
(b).}
\end{figure}

A more detailed analysis of the plot on Fig.~\ref{fig:Anticorrelation}
yields the coefficients of the linear combination with the best energy
resolution: $E=S2*0.715+S1*0.285$. Using this formula, an energy
resolution of 10.6\% FWHM was obtained at 122~keV for the whole fiducial
volume -- see Fig.~\ref{fig:EnergyRes}(a). For the central spot
with $R$$<$50~mm, where the effects from Compton scattering of
incoming $\gamma$\nobreakdash-rays in copper are minimal, the resolution
is 8.1\% FWHM and the two lines of the $^{57}$Co source are clearly
resolved as shown in Fig.~\ref{fig:EnergyRes}(b).

\section{Conclusions}

Position sensitivity is crucially important for a modern dark matter
detector as it allows one to drastically reduce the background by
considering only events inside an inner fiducial volume away from
any detector surfaces. A position-sensitive detector also offers better
energy resolution as it becomes possible to apply a position-dependent
correction to the energy. In the case of a scintillation detector,
the optimal performance of the position estimation algorithm depends
on how well the set of the PMT LRFs describes the detector response
to scintillation events. 

In the present work, a novel method for iterative reconstruction of
the light response functions from the calibration data acquired with
uncollimated $\gamma$\nobreakdash-ray source was developed and its
suitability has been proven for the real detector. Using the reconstructed
LRFs and applying the weighted least squares and maximum likelihood
methods to position and energy reconstruction, the performance of
the ZEPLIN-III detector was studied for 122~keV $\gamma$\nobreakdash-rays.
The measured performance for the inner part of the detector ($R$$<$100~mm)
is as follows: 
\begin{itemize}
\item spatial resolution of 13~mm FWHM in the horizontal plane for scintillation
signal (S1); 
\item spatial resolution of 1.6~mm FWHM for electro-luminescence signal
(S2);
\item energy resolution of 8.1\% FWHM for the combined (S1 and S2) signal. 
\end{itemize}
A more detailed description of the implementation of the position
reconstruction algorithms and their impact on the WIMP search with
ZEPLIN-III will be published as a separate paper. The developed method
can also be applied in scintillation cameras for medical imaging for
correction of non-uniformities and improving non-linearity associated
with both the scintillation crystal and the PMT array as well as those
due to the position reconstruction algorithm. The success of the new
method in mitigating significant performance irregularities suggests
that hardware components may be subject to less stringent requirements,
thereby reducing the cost of scintillation cameras. The method can
also be of advantage for regular quality control of gamma cameras.

\section{Acknowledgements }

The UK groups acknowledge the support of the Science \& Technology
Facilities Council (STFC) for the ZEPLIN\textendash{}III project and
for maintenance and operation of the underground Palmer laboratory
which is hosted by Cleveland Potash Ltd (CPL) at Boulby Mine, near
Whitby on the North-East coast of England. The project would not be
possible without the co-operation of the management and staff of CPL.
We also acknowledge support from a Joint International Project award,
held at ITEP and Imperial College, from the Russian Foundation of
Basic Research (08-02-91851 KO a) and the Royal Society. LIP\textendash{}Coimbra
acknowledges financial support from Fundação para a Ciência e Tecnologia
(FCT) through the project-grants CERN/FP/109320/2009, CERN/FP/116374/2010
and PTDC/FIS/67002/2006, as well as the postdoctoral grants SFRH/BPD/27054/2006,
SFRH/BPD/47320/2008 and SFRH/BPD/63096/2009. This work was supported
in part by SC Rosatom, contract \#H.4e.45.90.11.1059 from 10.03.2011.

\bibliographystyle{ieeetr}

\end{document}